\documentclass[twocolumn,superscriptaddress,prb]{revtex4-1}
\usepackage{graphicx}
\usepackage{natbib}
\usepackage{amsmath}

\begin{document}

\title{Optical far-field super-resolution microscopy using nitrogen vacancy center ensemble in bulk diamond}
\author{Shen Li}
\author{Xiang-dong Chen}
\email{xdch@ustc.edu.cn}
\author{Bo-Wen Zhao}
\author{Yang Dong}
\affiliation{Key Lab of Quantum Information, University of Science and
Technology of China, Hefei 230026, P.R. China}
\affiliation{Synergetic Innovation Center of Quantum Information $\&$ Quantum Physics, University of Science
and Technology of China, Hefei, 230026, China}
\author{Chong-Wen Zou}
\affiliation{National Synchrotron Radiation Laboratory, University of Science and Technology of China, Hefei, 230026, China}
\author{Guang-Can Guo}
\author{Fang-Wen Sun}
\affiliation{Key Lab of Quantum Information, University of Science and
Technology of China, Hefei 230026, P.R. China}
\affiliation{Synergetic Innovation Center of Quantum Information $\&$ Quantum Physics, University of Science
and Technology of China, Hefei, 230026, China}

\begin{abstract}
We demonstrate an optical far-field super-resolution microscopy using array of nitrogen vacancy centers in bulk diamond as near-field optical probes. The local optical field, which transmits through the nanostructures on the diamond surface, is measured by detecting the charge state conversion of nitrogen vacancy center. And the locating of nitrogen vacancy center with spatial resolution of 6.1 nm is realized with the charge state depletion nanoscopy. The nanostructures on the surface of diamond are then imaged with resolution below optical diffraction limit. The results offer an approach to built a general-purpose optical super-resolution microscopy and a convenient platform for high spatial resolution quantum sensing with nitrogen vacancy center.
\end{abstract}

\date{\today}

\maketitle

With stable fluorescence, optically initialized spin state and long spin coherence time, the nitrogen vacancy (NV) center in diamond is a promising candidate for quantum metrology\cite{Doherty2013phyrrevie,rondin2014rppreview}. High sensitive detection of electromagnetic field and temperature have been demonstrated with  NV center\cite{jel-nature2008-1,Vamivakas-nl2013,Jacques-science2014-domainwall,wra-nl2013-them,chang-2015nl-tem}, even in living biological cells\cite{lukin-nature-therm,Walsworth2016magsensing,lukin-2013naturecell}. Due to the sub-nanometer size, high spatial resolution is one of the most important advantage of NV based sensors. Typically, it is realized by scanning a specially prepared tip with attached nanodiamond\cite{Jacques-science2014-domainwall,Meriles-nc2015-tem,Vamivakas-nl2013,hollenberg-nl-15,Drezet2015-nvnsom}. However, the presence and movement of tips may change the local field distribution at the nanoscale \cite{localfield-2014np,Aksyuk-opticalmode-2016np}. Additionally, the preparation of the tips is usually difficult\cite{rsi-2016-tip}. An alternative method is to replace the scanning tips with high-density non-scanning array of NV centers, whose relative positions and statuses are obtained through optical far-field microscopy\cite{Awschalom-apl2010-magnetic,vecsen-njp,holl-2012-screp,holle2015magimaging}. However, the spatial resolution with conventional optical microscopy is limited by the diffraction. Recently, several different types of optical nanoscopy methods for NV center imaging have been developed\cite{adma2012sil,nl-2013-mw,wra2014PNASstorm,lukin-farfield,hell-nl-micro,chen201501,pengxi-2014-rcs}. Therefore, it is possible to use NV center ensemble arrays as probes to realize sub-diffraction spatial resolution quantum sensing.

\begin{figure}
  \centering
  \includegraphics[width=7cm]{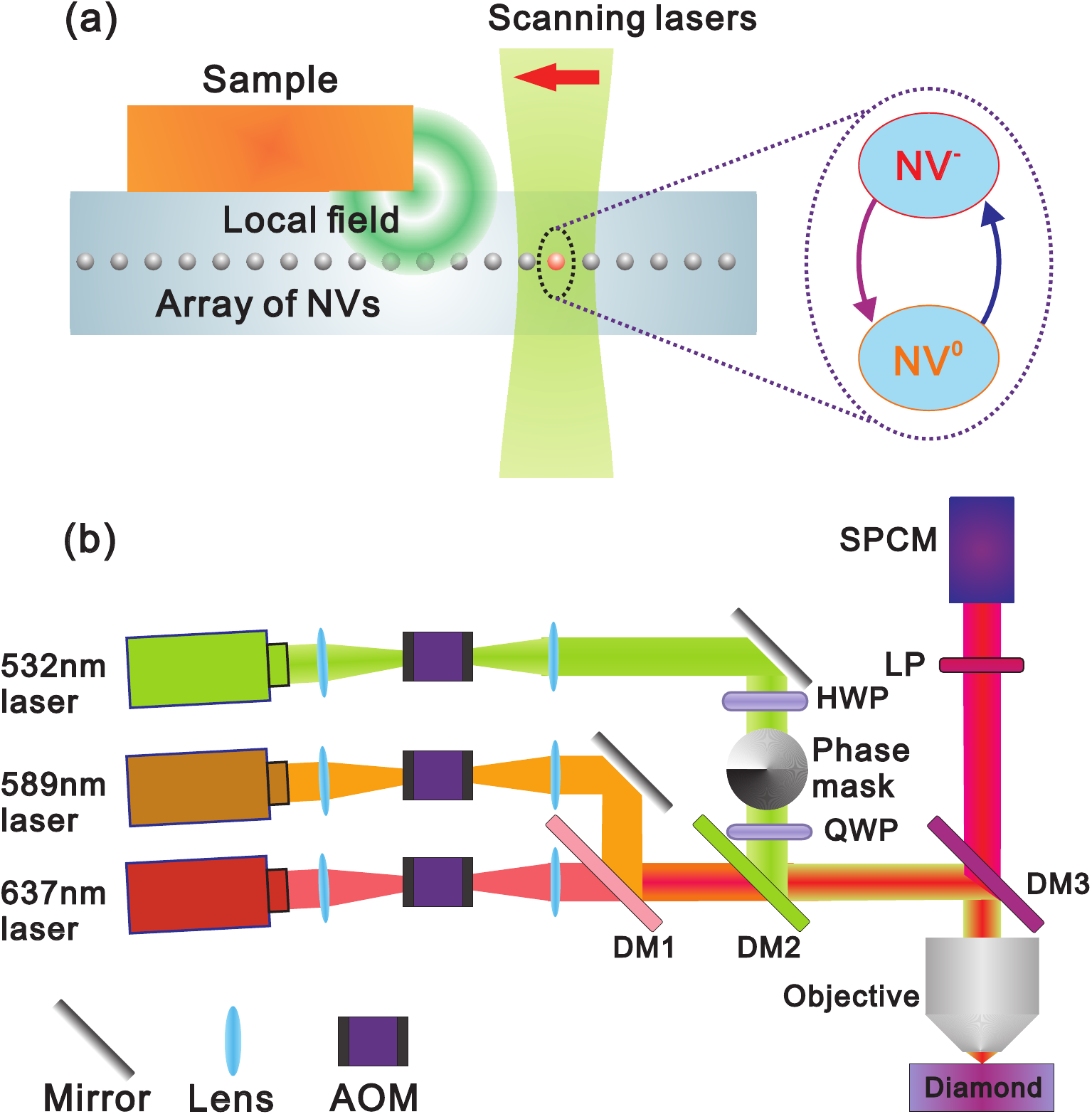}\\
  \caption{ (a) Illustration of local field detection with MANP. (b) The experimental setup for CSD nanoscopy of the NV center. The pulses of the laser beams are controlled by acousto-optic modulators (AOMs). The doughnut-shaped laser beam is generated by a vortex phase mask, and the polarization is controlled with a half-wave plate (HWP) and a quarter-wave plate (QWP). LP, long-pass filter with an edge wavelength of 668.9 nm. DM1-3 are DMs with edge wavelengths of 605 nm, 536.8 nm, and 658.8 nm, respectively. }\label{figsetup}\label{figprin}
\end{figure}

In particular, the local optical field detection with high spatial resolution is interesting in diverse fields ranging from nanophotonics to biosensing \cite{localfield-2014np,Vamivakas-nl2013,Aksyuk-opticalmode-2016np}. In this work, we demonstrate the local optical field detection with sub-diffraction resolution using NV center ensemble in bulk diamond as probes, as shown in Fig. \ref{figsetup}(a). The light transmitted through nanostructure on the diamond surface is measured with the charge state conversion of near-field NV center probes. And the high spatial resolution imaging of NV is realized with charge state depletion (CSD) nanoscopy. Subsequently, the sub-diffraction images of the nanostructures on the diamond plate surface are obtained via local optical field detection. A general-purpose super-resolution Microscopy with Array of Near-field Probes (MANP) is constructed based on the results. Furthermore, because the NV center is also sensitive to electromagnetic fields, temperature and pressure\cite{Doherty2014prlpressure}, the present technique provides a convenient method to establish a universal, nanoscale, multi-functional detection platform.

The NV center in diamond consists of a substitutional nitrogen atom and an adjacent vacancy and typically has two charge states: the neutral NV$^{0}$ and the negative NV$^{-}$, with zero phonon lines at 575 and 637 nm, respectively.  When a long-pass optical filter (edge wavelength 668.9 nm in this experiment) is used for the fluorescence detection, NV$^{0}$ is treated as the optical dark state, and  NV$^{-}$ is the bright state.
The conversion between the two states can be pumped with a wide spectral range of photons \cite{ioni-arxiv-2012,chen20132,hell-nl-micro}. The highest NV$^{-}$ and NV$^{0}$ populations are obtained with 532 nm and 637 nm lasers, respectively \cite{ioni-arxiv-2012,chen20132,chen20152prb}. The charge state conversion process of the NV center pumped by photon in visible region has been demonstrated to be a two-photon process involving a real excited state\cite{wra-lowtem-2012}. The charge state conversion rate quadratically increases as the optical field intensity increases in the visible region \cite{wra-lowtem-2012,ioni-arxiv-2012,chen20132}. Therefore, by detecting the charge state populations of NV centers, the optical field intensity can be measured.

\begin{figure}[ht]
  \centering
  \includegraphics[width=8.5cm]{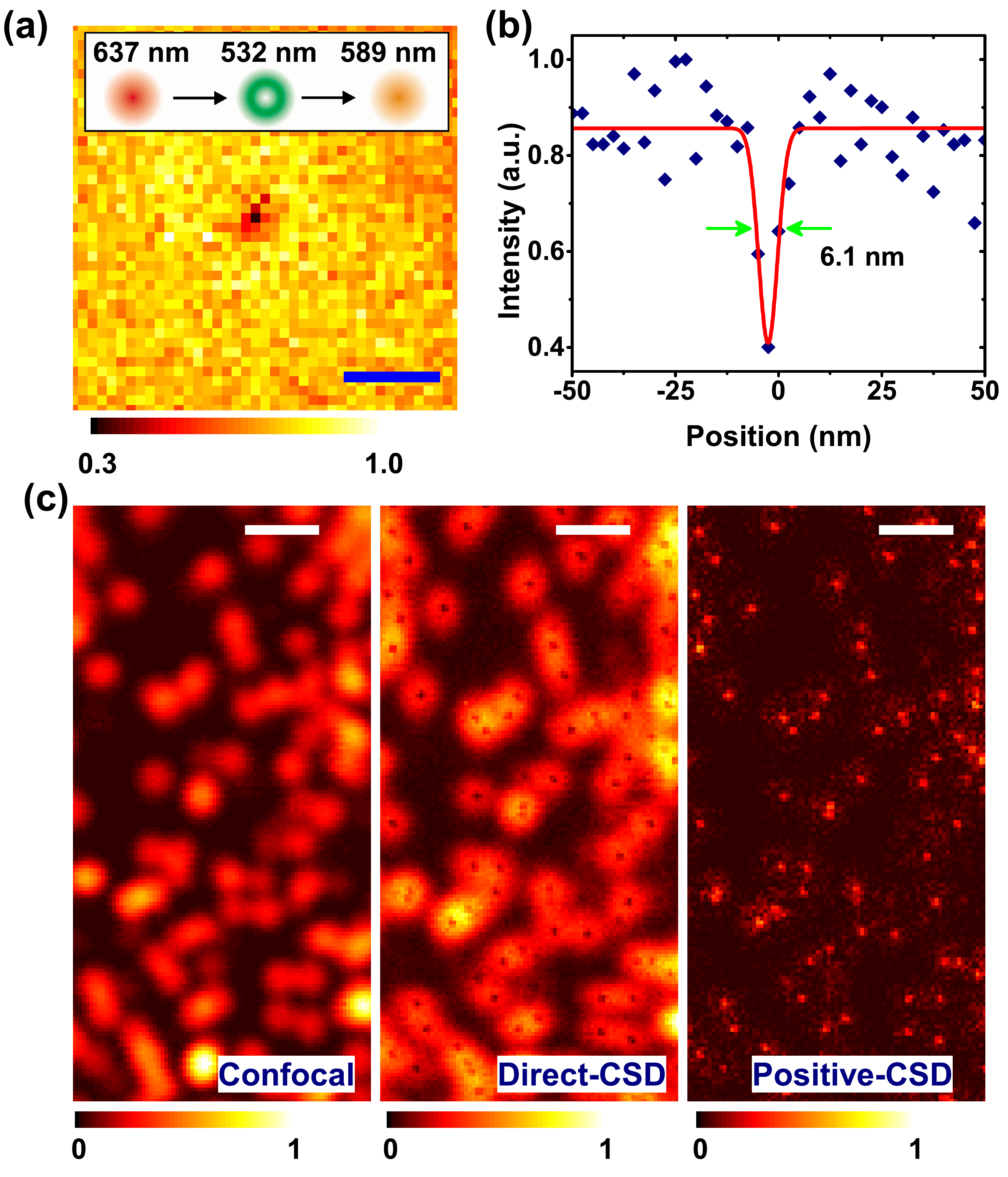}\\
  \caption{(a) CSD nanoscopy image with resolution of 6.1 nm. 532 nm doughnut-shaped depletion laser: 22 mW, 100 $\mu s$ duration. Insert is the laser pulse sequence. (b) Cross-section of the image in (a). (c) The positive-CSD image is obtained by subtracting the direct-CSD image from the confocal image. Pixel dwell time: 50 ms. 532 nm doughnut-shaped depletion laser beam: 80$\mu s$ duration, 0.13 mW average power.}\label{figbestreso}
\end{figure}

To achieve super-resolution imaging and sensing with MANP, locating the NV center with high spatial resolution is key. Here, CSD nanoscopy of the NV center is utilized for this purpose \cite{chen201501,hell-nl-micro}. The experimental setup is shown in Fig. \ref{figsetup}(b). A home-built scanning confocal microscope with a 0.95 numerical aperture objective was used to pump and image the NV center. A diamond plate with NV center inside was mounted on a piezo scanner. Different laser beams were aligned and combined using dichroic mirrors (DMs) and focused on the NV centers. The fluorescence of the NV$^{-}$ center was detected by a single photon counting module (SPCM).

For CSD nanoscopy used in our experiments, the NV center is firstly initialized to the NV$^{0}$ charge state with a 637 nm laser. Then, a doughnut-shaped 532 nm depletion laser beam is applied to convert the NV$^{0}$ charge state to NV$^{-}$ with a power-dependent rate. Finally, the NV$^{-}$ charge state population is detected with a 589 nm laser \cite{ioni-arxiv-2012,chen20132}. Only the photons emitted during the 589 nm laser pumping are counted. The initialization-depletion-detection procedure repeated hundreds of times to obtain sufficient photons for each scanning pixel. The resolution of CSD nanoscopy is determined by the depletion rate, which can be improved, in principle, by increasing the power of the doughnut-shaped beam \cite{chen201501,hell-nl-micro}.

\begin{figure}[ht]
  \centering
  \includegraphics[width=8cm]{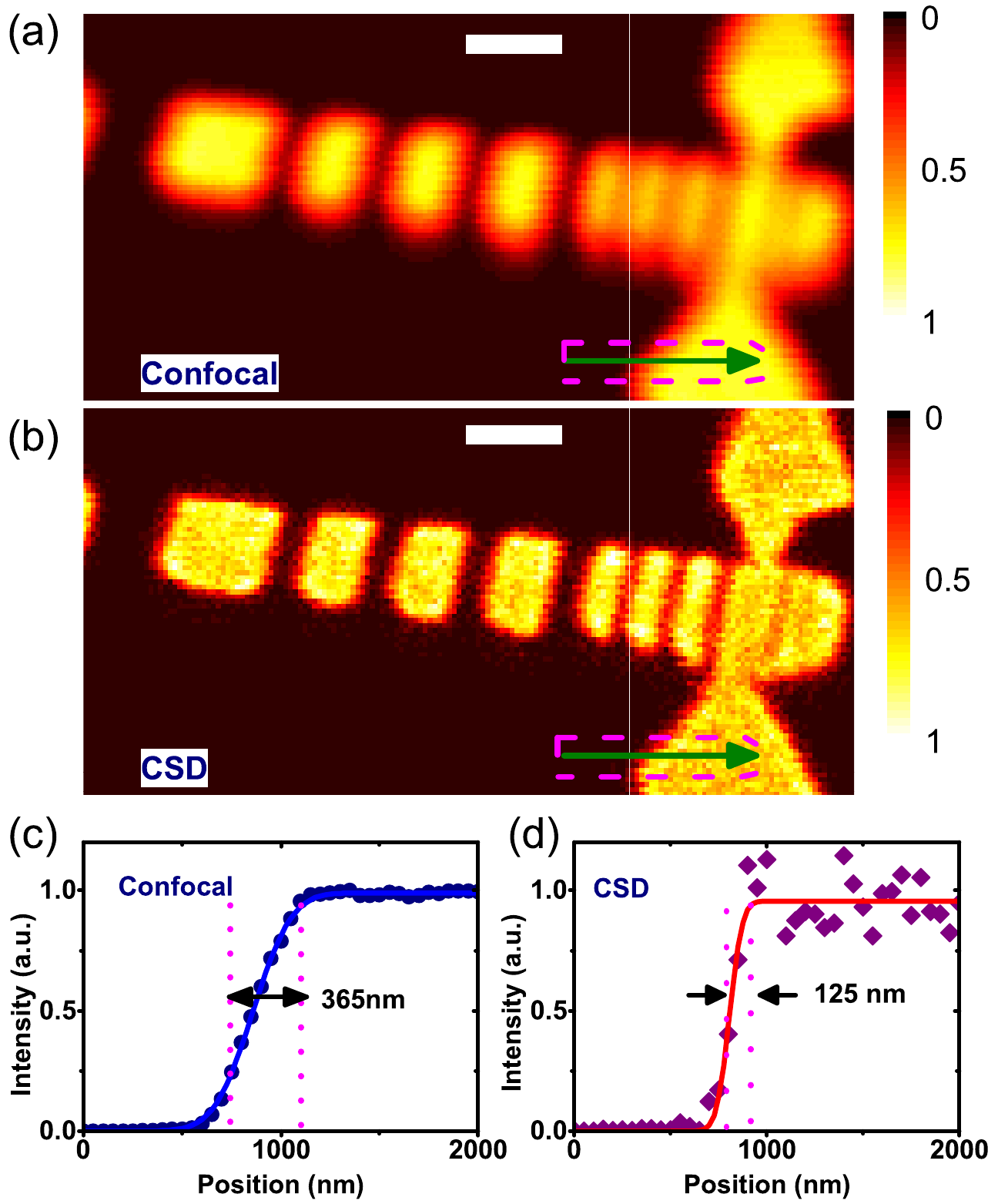}\\
  \caption{(a)(b) The images of NV center ensemble with confocal and CSD microscopy, respectively. (c)(d) The cross-section profiles corresponding to the arrows in (a)(b). Pixel dwell time: 50 ms. The 532 nm depletion beam: 0.3 mW average power, 100 $\mu s$ duration.}\label{csdensemble}
\end{figure}

Sub-10-nm-resolution CSD nanoscopy is achieved with high power doughnut-shaped depletion laser. An optical resolution of 6.1 nm is shown in Fig. \ref{figbestreso}(a)(b), which corresponds to approximately $1/56$ of the diffraction limit (1.22$\lambda/2N.A. \approx$ 341 nm for 532 nm laser beam). Here, the charge state depletion laser is 22 mW 532 nm doughnut-shaped laser beam with duration of 100$\mu s$ in each initialization-depletion-detection cycle. Meanwhile, the fluorescence of NV$^{-}$ is detected by a 0.1 mW 589 nm laser pulse with duration of 10 $\mu s$. The duty ratio of detection laser pulse is lower than 0.1 here. The signal to noise ratio (SNR), defined as the ratio of signal value to standard deviation of background, is approximate 6 in Fig. \ref{figbestreso}(a)(b). The SNR of CSD nanoscopy is lower than that of confocal microscopy. This is one of the major drawback for super-resolution microscopy method. In future, the charge state conversion rate of depletion laser should be improved to decrease the duration of depletion laser, and subsequently to increase the duty ratio of fluorescence detection laser.

The direct result of the 532 nm doughnut-shaped laser pumped CSD nanoscopy produces a negative image of the NV center, which is not convenient for the imaging of high-density NV centers. Experimentally, the positive images can be obtained by subtracting the signals of the direct-CSD images from the confocal images, as shown in Fig. \ref{figbestreso}(c). This positive-CSD image can then be used to detect the NV center ensemble in the following MANP experiments.

For MANP imaging, the distances between NV centers can be seen as the scanning step of the scanning tip microscopy. Therefore, high density NV center ensemble is needed for high resolution imaging. In the experiments, the NV center ensembles are produced by high dosage ($10^{13} cm^{-2}$) nitrogen ion implantation. However, nitrogen defect, as donor of electron\cite{Mizuochi-2016prb,apl-2016-density,Raymond-2009prb}, could change the level structure in diamond\cite{WRA,collins-2002level}. Therefore, high dosage ion implantation might change the charge state conversion process of NV center, which will subsequently affect the results of CSD nanoscopy. In Fig. \ref{csdensemble}, we show the CSD nanoscopy of NV center ensemble. The fluorescence of NV center ensemble is treated as the integration of signals from NV centers at different positions. The cross-sections in Fig. \ref{csdensemble}(c)(d) are fitted by the Gaussian model\cite{Bettiol-nc-15}:
\begin{equation}\label{distribution}
  I(x)\propto \int_{a}^{b} exp(-\frac{(x-x_{0})^{2}}{2\sigma^{2}})/\sigma dx_{0},
\end{equation}
where $a$ and $b$ are the edges of NV center ensemble. $exp(-\frac{(x-x_{0})^{2}}{2\sigma^{2}})/\sigma$ denotes the fluorescence point spread function of NV center at position $x_{0}$. Therefore, the resolution of microscopy is given by the full width at half maximum $2 \sqrt{2 ln2} \sigma$. Fitting the scanning-profile of CSD nanoscopy with the Eq.\ref{distribution}, the spatial resolution of CSD microscopy for NV ensemble in Fig. \ref{csdensemble} is 125 nm, which is below the diffraction limit. The results demonstrated that CSD nanoscopy is effective to NV center emsemble. A relative low power depletion laser is chosen for NV ensemble CSD imaging here. This is because the high intensity fluorescence emission of NV ensembles pumped by high power 532 nm depletion laser could saturate the SPCM, though the photons emitted during depletion process are not counted by the data acquisition card.

\begin{figure*}[ht]
  \centering
  \includegraphics[width=11cm]{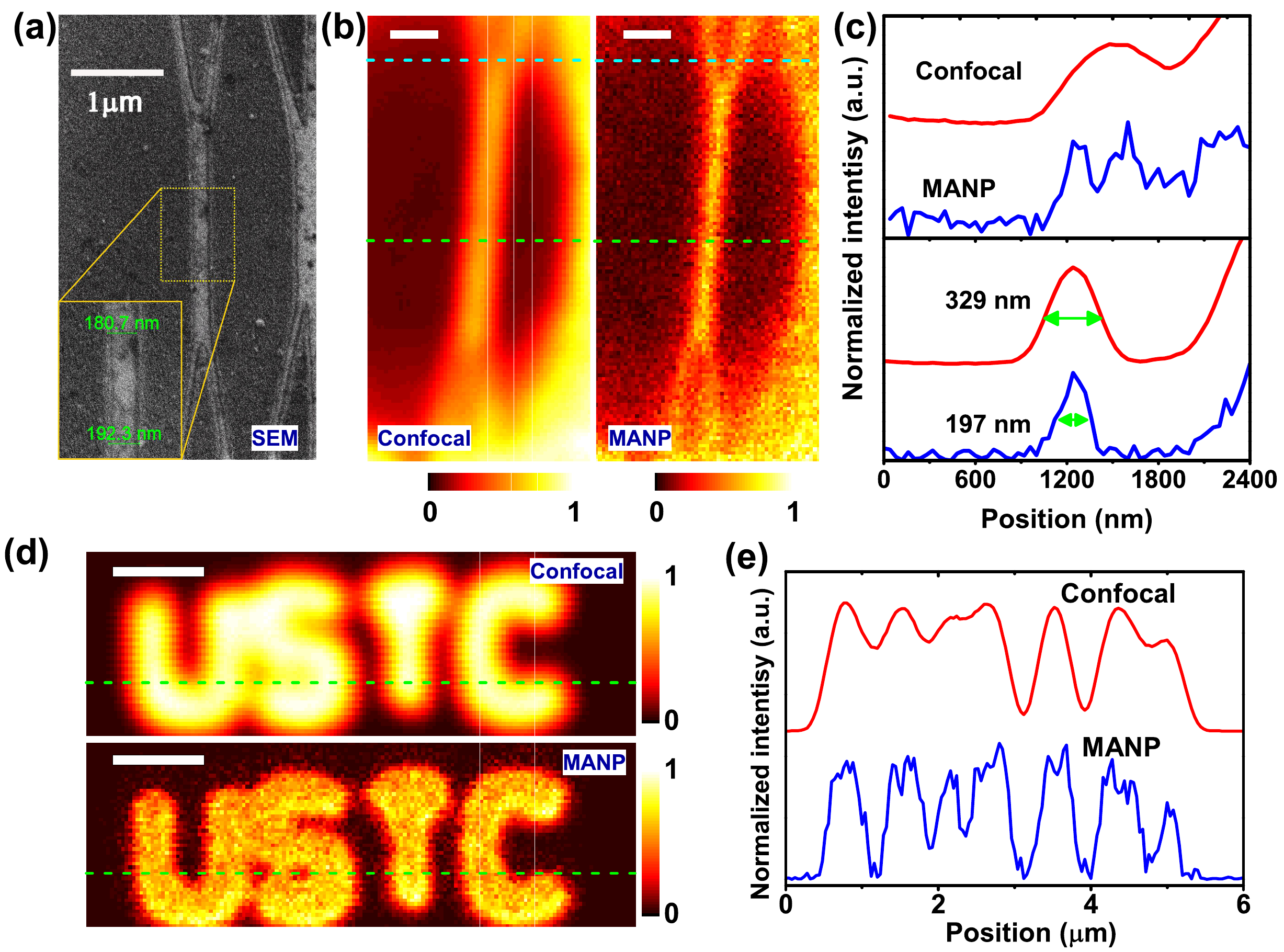}\\
  \caption{(a) SEM image of the titanium/aluminum film on diamond plate surface. (b) Optical confocal and MANP images of the titanium/aluminum film. (c) Cross-sections of the corresponding dashed lines in (b). (d) The confocal and MANP images of a chromium film. (e) Cross-section profiles of the dashed lines in (d). Pixel dwell time: 50 ms. 532 nm doughnut-shaped laser beam for MANP: 0.13 mW average power, 80 $\mu$s duration.}\label{figalium}
\end{figure*}

To demonstrate the application of MANP using NV center ensemble, we detect the local optical field of light transmitted through a metal nanostructure on the surface of a diamond plate. The optical field probes, which are arrays of NV centers, are produced by high dosage ion implantation prior to metal film deposition. The ion implantation energy is 15 keV. The mean depth of the NV centers is $20\pm7$ nm as estimated using SRIM\cite{srim}. The distance between the optical probe and the nanostructure is much shorter than the wavelength of the photons used in the experiment and is sufficient to detect the samples' optical near-field properties.
For local optical field detection, the same doughnut-shaped laser beam used for CSD nanoscopy is transmitted through the nanostructure to pump the NV center. Subsequently, the charge state conversion of the NV center is affected by the shape of the nanostructure. Therefore, by detecting the fluorescence intensity of the NV center via CSD nanoscopy, images of samples with sub-diffraction-limit resolution can be obtained.

To quantify the spatial resolution of MANP, we present the scanning electron microscope (SEM) and optical images of a titanium/aluminum (5/15 nm thickness) nanostructure in Fig. \ref{figalium}(a)(b). The confocal images are obtained by directly detecting the fluorescence of the NV center with a single Gaussian-shaped laser beam. The positive images of MANP with the 532 nm doughnut beam are collected using the method depicted in Fig. \ref{figbestreso}(c). The bright areas of the optical images indicate the absence of the metal film. The fork structures, which cannot be distinguished in the confocal image, are clearly identified by MANP. The width of the slot measured by SEM is approximately 190 nm. By fitting the cross-section in Fig. \ref{figalium}(c) with Eq.(\ref{distribution}), the resolution of MANP in Fig. \ref{figalium}(b) is estimated to be 197 nm. Thus, the resolution achieved when imaging this nanostructure with MANP is below the diffraction limit.

Sub-diffraction-limit MANP can be effectively applied to various materials and patterns. Fig. \ref{figalium}(d)(e) show the imaging of the English letters 'USTC' made of chromium film (22 nm thickness). The spatial resolution is also significantly improved by MANP. This result demonstrates the generalizability of the NV-center-based super-resolution MANP. Because the band gap of diamond is 5.5 eV and because the charge state conversion can be pumped with a wide range of wavelengths \cite{chen20132}, MANP with NV centers can be applied to detect local optical fields in the spectral range from near ultraviolet to near infrared.

Like CSD nanoscopy, the resolution of MANP can be improved by increasing the power of doughnut-shaped laser beam, and is theoretically unlimited. Technically, high laser power would induce heating of sample, or photodamage to biological cell. And imperfect shape of the depletion laser will limit the best spatial resolution. Besides super-resolution imaging, sub-diffraction-limit MANP with CSD nanoscopy on the NV center can also be used for nanophotonics investigations, such as scattering and surface plasmon resonance of optical nanoanttenna\cite{lukin-2011prl,Quidant-nl2014,Vamivakas-nl2013}. Furthermore, since quantum imaging has been realized based on electron spin manipulation and diffraction-limited optical microscopy of NV center ensemble\cite{holle2015magimaging,holl-2012-screp,lukin-2013naturecell}, NV-center-based super-resolution MANP can be applied to sense electromagnetic fields and temperature with high spatial resolution and detection sensitivity. Compared with the method of using scanning nanodiamond tips \cite{Vamivakas-nl2013,Meriles-nc2015-tem,Jacques-science2014-domainwall}, nanoscale sensing with MANP is much more robust as the scanning tips are not needed.

In conclusion, we developed a universal, super-resolution, optical far-field microscope based on local optical field sensing with the NV center ensemble in bulk diamond. With the sub-10 nm spatial resolving capability of the NV center, the nanostructures of different metal films on a diamond surface were imaged with resolutions below the diffraction limit. The system of MANP has the potential to be developed to a multi-functional super-resolution microscope with high detection sensitivity, which will help to explore quantum sensing at the nanoscale.

This work was supported by the Strategic Priority Research Program(B) of the
Chinese Academy of Sciences (Grant No. XDB01030200), the National Natural
Science Foundation of China (Nos. 11374290, 91536219, 61522508, 11504363),
the Fundamental Research Funds for the Central Universities, the China
Postdoctoral Science Foundation (No.2016T90565).

\end{document}